\documentclass[a4paper,11pt]{article}
\usepackage{pos}

\usepackage{graphicx}
\usepackage{amsmath}
\usepackage[dvipsnames]{xcolor}
\usepackage{hyperref}
\usepackage{cleveref}
\usepackage{booktabs}
\usepackage{physics}
\usepackage{dsfont}
\usepackage{tcolorbox}
\usepackage{enumitem}
\usepackage{graphicx}
\usepackage{subfig}
\usepackage{wrapfig}
\usepackage[export]{adjustbox}
\usepackage{setspace}
\usepackage{bbm}
\usepackage{bm}
\usepackage{orcidlink}
\usepackage{soul}
\usepackage{subcaption}
\usepackage{array}
\usepackage{multirow}
\usepackage[table]{xcolor}
\usepackage{caption}

\usepackage{tikz}
\usepackage[compat = 1.1.0]{tikz-feynman}
\usepackage[customcolors]{hf-tikz}
\usetikzlibrary{positioning,calc,intersections,patterns.meta}

\definecolor{llbl}{HTML}{DBE5F1}

\let\OLDthebibliography\thebibliography
\renewcommand\thebibliography[1]{
  \OLDthebibliography{#1}
  \setlength{\parskip}{1pt}
  \setlength{\itemsep}{0.4ex plus 0.2ex}
}

\newcommand{\smaller}[1]{{\mathchoice
  {\scalebox{1.1}{$\scriptstyle #1$}}
  {\scalebox{1.1}{$\scriptstyle #1$}}
  {\scalebox{1.1}{$\scriptscriptstyle #1$}}
  {\scalebox{1.1}{$\scriptscriptstyle #1$}}
}}

\newcommand{\EM}[0]{\smaller{\rm EM}}
\newcommand{\QCD}[0]{\smaller{\rm QCD}}
\newcommand{\QCED}[0]{\smaller{\rm QCD+QED}}
\newcommand{\EXP}[0]{\smaller{\rm exp}}
\newcommand{\IR}[0]{\smaller{\rm IR}}
\newcommand{\IB}[0]{\smaller{\rm IB}}
\newcommand{\W}[0]{\smaller{W}}
\newcommand{\ph}[0]{{\phantom{1}}}

\tikzset{
  squaredot/.style={
    fill,
    draw=black,
    minimum size=4pt,
    inner sep=0pt,
    rectangle
  }
}

\newlength{\QCDlen}
\newlength{\QCDvertA}
\newlength{\QCDvertB}
\newlength{\QCDvertC}

\NewDocumentCommand\semiloop{O{black}mmmO{}O{above}}
{%
\draw[#1] let \p1 = ($(#3)-(#2)$) in
  (#3) arc (#4:{#4+180}:{0.5*veclen(\x1,\y1)}) node[midway, #6] {#5};
}

\newcommand{\Sestr}{
\begin{tikzpicture}[
dot/.style={draw,circle,minimum size=2.5pt,inner sep=0pt,outer sep=0pt,fill=Green, color=Green}
]

\begin{feynman}
\vertex (n1) ;
\vertex[below=0.32*\QCDvertA of n1] (n11)    ;
\vertex[below=0.64*\QCDvertA of n1] (n12)    ;
\vertex[right=\QCDlen of n1] (n2)   ;
\vertex[right=\QCDlen of n11] (n21) ;
\vertex[right=\QCDlen of n12] (n22) ;

\vertex at ($(n1)!0.95!(n2) + (0, \QCDvertB)$) (n3) ;
\vertex at ($(n1)!0.95!(n2) + (0, \QCDvertC)$) (n4) ;

\vertex at ($(n1)!0.50!(n2)$) (m1);
\vertex[above=1.2*\QCDvertA of m1] (m2);

\vertex[dot] at ($(n12)!0.25!(n22)$) (c1);
\vertex[dot] at ($(n12)!0.75!(n22)$) (c2);

    \diagram*{
        (n1)  -- [fermion, color=Red   ] (m1)  ,
        (n11) -- [fermion, color=Red   ] (n21)  , 
        (m1)  -- [fermion, color=Apricot] (n2)  ,
        (n12) -- [fermion, color=black  ] (n22)  , 
        (n3) -- [fermion, color=Blue] (m2) , 
        (m2) -- [fermion, color=Blue] (n4) , 
        (c1) -- [photon, thick, in =-120, out =-60, color=Green] (c2) ,
    };

\node[dot] at (c1);
\node[dot] at (c2);
\node[dot, white] at ($(n1) - (0, {1.30*\QCDvertA})$);

\draw [color=gray!60, fill=gray!25, ] (n11) ellipse ({0.40*\QCDvertA} and {0.70*\QCDvertA});
\draw [color=gray!60, fill=gray!25, ] (n21) ellipse ({0.40*\QCDvertA} and {0.70*\QCDvertA});
\node at (n11) {$\Sigma$};
\node at (n21) {$\Lambda$};
\node[squaredot] at (m1);
\node[squaredot] at (m2);
\node[color=Blue] at ($(n4) + (0.25, 0)$) {\(e\)};
\node[color=Blue] at ($(n3) + (0.25, 0)$) {\(\bar{\nu}\)};

\end{feynman}
\end{tikzpicture}
}

\newcommand{\Seup}{
\begin{tikzpicture}[
dot/.style={draw,circle,minimum size=2.5pt,inner sep=0pt,outer sep=0pt,fill=Green, color=Green}
]

\begin{feynman}
\vertex (n1) ;
\vertex[below=0.32*\QCDvertA of n1] (n11)    ;
\vertex[below=0.64*\QCDvertA of n1] (n12)    ;
\vertex[right=\QCDlen of n1] (n2)   ;
\vertex[right=\QCDlen of n11] (n21) ;
\vertex[right=\QCDlen of n12] (n22) ;

\vertex at ($(n1)!0.95!(n2) + (0, \QCDvertB)$) (n3) ;
\vertex at ($(n1)!0.95!(n2) + (0, \QCDvertC)$) (n4) ;

\vertex at ($(n1)!0.50!(n2)$) (m1);
\vertex[above=1.2*\QCDvertA of m1] (m2);

\vertex[dot] at ($(n1)!0.25!(m1)$) (c1);
\vertex[dot] at ($(n1)!0.75!(m1)$) (c2);

    \diagram*{
        (n1)  -- [fermion, color=Red   ] (m1)  ,
        (n11) -- [fermion, color=Red   ] (n21)  , 
        (m1)  -- [fermion, color=Apricot] (n2)  ,
        (n12) -- [fermion, color=black  ] (n22)  , 
        (n3) -- [fermion, color=Blue] (m2) , 
        (m2) -- [fermion, color=Blue] (n4) , 
        (c1) -- [photon, thick, in =120, out =60, color=Green] (c2) ,
    };

\node[dot] at (c1);
\node[dot] at (c2);
\node[dot, white] at ($(n1) - (0, {1.30*\QCDvertA})$);

\draw [color=gray!60, fill=gray!25, ] (n11) ellipse ({0.40*\QCDvertA} and {0.70*\QCDvertA});
\draw [color=gray!60, fill=gray!25, ] (n21) ellipse ({0.40*\QCDvertA} and {0.70*\QCDvertA});
\node at (n11) {$\Sigma$};
\node at (n21) {$\Lambda$};
\node[squaredot] at (m1);
\node[squaredot] at (m2);
\node[color=Blue] at ($(n4) + (0.25, 0)$) {\(e\)};
\node[color=Blue] at ($(n3) + (0.25, 0)$) {\(\bar{\nu}\)};

\end{feynman}
\end{tikzpicture}
}

\newcommand{\Selep}{
\begin{tikzpicture}[
dot/.style={draw,circle,minimum size=2.5pt,inner sep=0pt,outer sep=0pt,fill=Green, color=Green}
]

\begin{feynman}
\vertex (n1) ;
\vertex[below=0.32*\QCDvertA of n1] (n11)    ;
\vertex[below=0.64*\QCDvertA of n1] (n12)    ;
\vertex[right=\QCDlen of n1] (n2)   ;
\vertex[right=\QCDlen of n11] (n21) ;
\vertex[right=\QCDlen of n12] (n22) ;

\vertex at ($(n1)!0.95!(n2) + (0, \QCDvertB)$) (n3) ;
\vertex at ($(n1)!0.95!(n2) + (0, \QCDvertC)$) (n4) ;

\vertex at ($(n1)!0.50!(n2)$) (m1);
\vertex[above=1.2*\QCDvertA of m1] (m2);

\vertex[dot] at ($(m2)!0.2!(n4)$) (c1);
\vertex[dot] at ($(m2)!0.8!(n4)$) (c2);

    \diagram*{
        (n1)  -- [fermion, color=Red   ] (m1)  ,
        (n11) -- [fermion, color=Red   ] (n21)  , 
        (m1)  -- [fermion, color=Apricot] (n2)  ,
        (n12) -- [fermion, color=black  ] (n22)  , 
        (n3) -- [fermion, color=Blue] (m2) , 
        (m2) -- [fermion, color=Blue] (n4) , 
        (c1) -- [photon, thick, in =-120, out =-60, color=Green] (c2) ,
    };

\node[dot] at (c1);
\node[dot] at (c2);
\node[dot, white] at ($(n1) - (0, {1.30*\QCDvertA})$);

\draw [color=gray!60, fill=gray!25, ] (n11) ellipse ({0.40*\QCDvertA} and {0.70*\QCDvertA});
\draw [color=gray!60, fill=gray!25, ] (n21) ellipse ({0.40*\QCDvertA} and {0.70*\QCDvertA});
\node at (n11) {$\Sigma$};
\node at (n21) {$\Lambda$};
\node[squaredot] at (m1);
\node[squaredot] at (m2);
\node[color=Blue] at ($(n4) + (0.25, 0)$) {\(e\)};
\node[color=Blue] at ($(n3) + (0.25, 0)$) {\(\bar{\nu}\)};

\end{feynman}
\end{tikzpicture}
}

\newcommand{\Exup}{
\begin{tikzpicture}[
dot/.style={draw,circle,minimum size=2.5pt,inner sep=0pt,outer sep=0pt,fill=Green, color=Green}
]

\begin{feynman}
\vertex (n1) ;
\vertex[below=0.32*\QCDvertA of n1] (n11)    ;
\vertex[below=0.64*\QCDvertA of n1] (n12)    ;
\vertex[right=\QCDlen of n1] (n2)   ;
\vertex[right=\QCDlen of n11] (n21) ;
\vertex[right=\QCDlen of n12] (n22) ;

\vertex at ($(n1)!0.95!(n2) + (0, \QCDvertB)$) (n3) ;
\vertex at ($(n1)!0.95!(n2) + (0, \QCDvertC)$) (n4) ;

\vertex at ($(n1)!0.50!(n2)$) (m1);
\vertex[above=1.2*\QCDvertA of m1] (m2);

\vertex[dot] at ($(n1)!0.25!(m1)$) (c1);
\vertex[dot] at ($(n12)!0.80!(n22)$) (c2);

    \diagram*{
        (n1)  -- [fermion, color=Red   ] (m1)  ,
        (n11) -- [fermion, color=Red   ] (n21)  , 
        (m1)  -- [fermion, color=Apricot] (n2)  ,
        (n12) -- [fermion, color=black  ] (n22)  , 
        (n3) -- [fermion, color=Blue] (m2) , 
        (m2) -- [fermion, color=Blue] (n4) , 
        (c1) -- [photon, thick, bend right, color=Green] (c2) ,
    };

\node[dot] at (c1);
\node[dot] at (c2);
\node[dot, white] at ($(n1) - (0, {1.30*\QCDvertA})$);

\draw [color=gray!60, fill=gray!25, ] (n11) ellipse ({0.40*\QCDvertA} and {0.70*\QCDvertA});
\draw [color=gray!60, fill=gray!25, ] (n21) ellipse ({0.40*\QCDvertA} and {0.70*\QCDvertA});
\node at (n11) {$\Sigma$};
\node at (n21) {$\Lambda$};
\node[squaredot] at (m1);
\node[squaredot] at (m2);
\node[color=Blue] at ($(n4) + (0.25, 0)$) {\(e\)};
\node[color=Blue] at ($(n3) + (0.25, 0)$) {\(\bar{\nu}\)};

\end{feynman}
\end{tikzpicture}
}

\newcommand{\Qmstr}{
\begin{tikzpicture}[
dot/.style={draw,circle,minimum size=2.5pt,inner sep=0pt,outer sep=0pt,fill=Green, color=Green}
]

\begin{feynman}
\vertex (n1) ;
\vertex[below=0.32*\QCDvertA of n1] (n11)    ;
\vertex[below=0.64*\QCDvertA of n1] (n12)    ;
\vertex[right=\QCDlen of n1] (n2)   ;
\vertex[right=\QCDlen of n11] (n21) ;
\vertex[right=\QCDlen of n12] (n22) ;

\vertex at ($(n1)!0.95!(n2) + (0, \QCDvertB)$) (n3) ;
\vertex at ($(n1)!0.95!(n2) + (0, \QCDvertC)$) (n4) ;

\vertex at ($(n1)!0.50!(n2)$) (m1);
\vertex[above=1.2*\QCDvertA of m1] (m2);

\vertex[dot] at ($(n12)!0.25!(n22)$) (c1);

    \diagram*{
        (n1)  -- [fermion, color=Red   ] (m1)  ,
        (n11) -- [fermion, color=Red   ] (n21)  , 
        (m1)  -- [fermion, color=Apricot] (n2)  ,
        (n12) -- [fermion, color=black  ] (n22)  , 
        (n3) -- [fermion, color=Blue] (m2) , 
        (m2) -- [fermion, color=Blue] (n4) , 
    };
    
\node[rectangle, draw, minimum width=5pt, minimum height=5pt, inner sep=0pt,, color=Green, pattern={Hatch[angle=45,distance=1.2pt,xshift=.1pt]}, pattern color=Green!70!black] at (c1);
\node[dot, white] at ($(n1) - (0, {1.30*\QCDvertA})$);

\draw [color=gray!60, fill=gray!25, ] (n11) ellipse ({0.40*\QCDvertA} and {0.70*\QCDvertA});
\draw [color=gray!60, fill=gray!25, ] (n21) ellipse ({0.40*\QCDvertA} and {0.70*\QCDvertA});
\node at (n11) {$\Sigma$};
\node at (n21) {$\Lambda$};
\node[squaredot] at (m1);
\node[squaredot] at (m2);
\node[color=Blue] at ($(n4) + (0.25, 0)$) {\(e\)};
\node[color=Blue] at ($(n3) + (0.25, 0)$) {\(\bar{\nu}\)};

\end{feynman}
\end{tikzpicture}
}

\newcommand{\Qmup}{
\begin{tikzpicture}[
dot/.style={draw,circle,minimum size=2.5pt,inner sep=0pt,outer sep=0pt,fill=Green, color=Green}
]

\begin{feynman}
\vertex (n1) ;
\vertex[below=0.32*\QCDvertA of n1] (n11)    ;
\vertex[below=0.64*\QCDvertA of n1] (n12)    ;
\vertex[right=\QCDlen of n1] (n2)   ;
\vertex[right=\QCDlen of n11] (n21) ;
\vertex[right=\QCDlen of n12] (n22) ;

\vertex at ($(n1)!0.95!(n2) + (0, \QCDvertB)$) (n3) ;
\vertex at ($(n1)!0.95!(n2) + (0, \QCDvertC)$) (n4) ;

\vertex at ($(n1)!0.50!(n2)$) (m1);
\vertex[above=1.2*\QCDvertA of m1] (m2);

\vertex[dot] at ($(n11)!0.25!(n21)$) (c1);

    \diagram*{
        (n1)  -- [fermion, color=Red   ] (m1)  ,
        (n11) -- [fermion, color=Red   ] (n21)  , 
        (m1)  -- [fermion, color=Apricot] (n2)  ,
        (n12) -- [fermion, color=black  ] (n22)  , 
        (n3) -- [fermion, color=Blue] (m2) , 
        (m2) -- [fermion, color=Blue] (n4) , 
    };
    
\node[rectangle, draw, minimum width=5pt, minimum height=5pt, inner sep=0pt,, color=Green, pattern={Hatch[angle=45,distance=1.2pt,xshift=.1pt]}, pattern color=Green!70!black] at (c1);
\node[dot, white] at ($(n1) - (0, {1.30*\QCDvertA})$);

\draw [color=gray!60, fill=gray!25, ] (n11) ellipse ({0.40*\QCDvertA} and {0.70*\QCDvertA});
\draw [color=gray!60, fill=gray!25, ] (n21) ellipse ({0.40*\QCDvertA} and {0.70*\QCDvertA});
\node at (n11) {$\Sigma$};
\node at (n21) {$\Lambda$};
\node[squaredot] at (m1);
\node[squaredot] at (m2);
\node[color=Blue] at ($(n4) + (0.25, 0)$) {\(e\)};
\node[color=Blue] at ($(n3) + (0.25, 0)$) {\(\bar{\nu}\)};

\end{feynman}
\end{tikzpicture}
}

\newcommand{\Sea}{
\begin{tikzpicture}[
dot/.style={draw,circle,minimum size=2.5pt,inner sep=0pt,outer sep=0pt,fill=Green, color=Green}
]

\begin{feynman}
\vertex (n1) ;
\vertex[below=0.32*\QCDvertA of n1] (n11)    ;
\vertex[below=0.64*\QCDvertA of n1] (n12)    ;
\vertex[right=\QCDlen of n1] (n2)   ;
\vertex[right=\QCDlen of n11] (n21) ;
\vertex[right=\QCDlen of n12] (n22) ;

\vertex at ($(n1)!0.95!(n2) + (0, \QCDvertB)$) (n3) ;
\vertex at ($(n1)!0.95!(n2) + (0, \QCDvertC)$) (n4) ;

\vertex at ($(n1)!0.50!(n2)$) (m1);
\vertex[above=1.2*\QCDvertA of m1] (m2);

\vertex at ($(n1)!0.35!(n2) + (0, 0.7*\QCDvertA - 0.32*\QCDvertA)$) (c1);
\vertex at ($(n1)!0.35!(n2) + (0, 1.3*\QCDvertA - 0.32*\QCDvertA)$) (c2);

\vertex at ($(n1)!0.10!(n2) + (0, 0.7*\QCDvertA + 0.10cm)$) (c3);
\vertex at ($(n1)!0.10!(n2) + (0, 1.3*\QCDvertA + 0.10cm)$) (c4);
\vertex at ($(n1)!0.10!(n2) + (0.3*\QCDvertA, 1.0*\QCDvertA + 0.10cm)$) (c5);

    \diagram*{
        (n1)  -- [fermion, color=Red   ] (m1)  ,
        (n11) -- [fermion, color=Red   ] (n21)  , 
        (m1)  -- [fermion, color=Apricot] (n2)  ,
        (n12) -- [fermion, color=black  ] (n22)  , 
        (n3) -- [fermion, color=Blue] (m2) , 
        (m2) -- [fermion, color=Blue] (n4) , 
        (c1) -- [photon, thick, color=Green] (c2) ,
    };
    
\semiloop[]{c2}{c1}{-90}[][right];
\semiloop[fermion]{c1}{c2}{90};

\semiloop[]{c4}{c3}{-90}[][right];
\semiloop[fermion]{c3}{c4}{90};

\node[dot] at (c1);
\node[dot] at (c2);
\node[dot, white] at ($(n1) - (0, {1.30*\QCDvertA})$);

\node[rectangle, draw, minimum width=5pt, minimum height=5pt, inner sep=0pt,, color=Green, pattern={Hatch[angle=45,distance=1.2pt,xshift=.1pt]}, pattern color=Green!70!black] at (c5);

\draw [color=gray!60, fill=gray!25, ] (n11) ellipse ({0.40*\QCDvertA} and {0.70*\QCDvertA});
\draw [color=gray!60, fill=gray!25, ] (n21) ellipse ({0.40*\QCDvertA} and {0.70*\QCDvertA});
\node at (n11) {$\Sigma$};
\node at (n21) {$\Lambda$};
\node[squaredot] at (m1);
\node[squaredot] at (m2);
\node[color=Blue] at ($(n4) + (0.25, 0)$) {\(e\)};
\node[color=Blue] at ($(n3) + (0.25, 0)$) {\(\bar{\nu}\)};
\node at ($(n1)!0.23!(n2) + (0, 1.1*\QCDvertA - 0.15)$) {\(\Big/\)};

\end{feynman}
\end{tikzpicture}}

\newcommand{\Sealep}{
\begin{tikzpicture}[
dot/.style={draw,circle,minimum size=2.5pt,inner sep=0pt,outer sep=0pt,fill=Green, color=Green}
]

\begin{feynman}
\vertex (n1) ;
\vertex[below=0.32*\QCDvertA of n1] (n11)    ;
\vertex[below=0.64*\QCDvertA of n1] (n12)    ;
\vertex[right=\QCDlen of n1] (n2)   ;
\vertex[right=\QCDlen of n11] (n21) ;
\vertex[right=\QCDlen of n12] (n22) ;

\vertex at ($(n1)!0.95!(n2) + (0, \QCDvertB)$) (n3) ;
\vertex at ($(n1)!0.95!(n2) + (0, \QCDvertC)$) (n4) ;

\vertex at ($(n1)!0.50!(n2)$) (m1);
\vertex[above=1.2*\QCDvertA of m1] (m2);

\vertex[dot] at ($(m2)!0.80!(n4)$) (c1);

\vertex at ($(n1)!0.200!(n2) + (0, 0.7*\QCDvertA)$) (c2);
\vertex at ($(n1)!0.200!(n2) + (0, 1.3*\QCDvertA)$) (c3);

    \diagram*{
        (n1)  -- [fermion, color=Red   ] (m1)  ,
        (n11) -- [fermion, color=Red   ] (n21)  , 
        (m1)  -- [fermion, color=Apricot] (n2)  ,
        (n12) -- [fermion, color=black  ] (n22)  , 
        (n3) -- [fermion, color=Blue] (m2) , 
        (m2) -- [fermion, color=Blue] (n4) , 
        (c1) -- [photon, thick, bend left, color=Green] (c2) ,
    };
    
\semiloop[]{c3}{c2}{-90}[][right];
\semiloop[fermion]{c2}{c3}{90};

\node[dot] at (c1);
\node[dot] at (c2);
\node[dot, white] at ($(n1) - (0, {1.30*\QCDvertA})$);

\draw [color=gray!60, fill=gray!25, ] (n11) ellipse ({0.40*\QCDvertA} and {0.70*\QCDvertA});
\draw [color=gray!60, fill=gray!25, ] (n21) ellipse ({0.40*\QCDvertA} and {0.70*\QCDvertA});
\node at (n11) {$\Sigma$};
\node at (n21) {$\Lambda$};
\node[squaredot] at (m1);
\node[squaredot] at (m2);
\node[color=Blue] at ($(n4) + (0.25, 0)$) {\(e\)};
\node[color=Blue] at ($(n3) + (0.25, 0)$) {\(\bar{\nu}\)};

\end{feynman}
\end{tikzpicture}}

\newcommand{\Seaval}{
\begin{tikzpicture}[
dot/.style={draw,circle,minimum size=2.5pt,inner sep=0pt,outer sep=0pt,fill=Green, color=Green}
]

\begin{feynman}
\vertex (n1) ;
\vertex[below=0.32*\QCDvertA of n1] (n11)    ;
\vertex[below=0.64*\QCDvertA of n1] (n12)    ;
\vertex[right=\QCDlen of n1] (n2)   ;
\vertex[right=\QCDlen of n11] (n21) ;
\vertex[right=\QCDlen of n12] (n22) ;

\vertex at ($(n1)!0.95!(n2) + (0, \QCDvertB)$) (n3) ;
\vertex at ($(n1)!0.95!(n2) + (0, \QCDvertC)$) (n4) ;

\vertex at ($(n1)!0.50!(n2)$) (m1);
\vertex[above=1.2*\QCDvertA of m1] (m2);

\vertex[dot] at ($(n1)!0.80!(m1)$) (c1);

\vertex at ($(n1)!0.200!(n2) + (0, 0.7*\QCDvertA)$) (c2);
\vertex at ($(n1)!0.200!(n2) + (0, 1.3*\QCDvertA)$) (c3);

    \diagram*{
        (n1)  -- [fermion, color=Red   ] (m1)  ,
        (n11) -- [fermion, color=Red   ] (n21)  , 
        (m1)  -- [fermion, color=Apricot] (n2)  ,
        (n12) -- [fermion, color=black  ] (n22)  , 
        (n3) -- [fermion, color=Blue] (m2) , 
        (m2) -- [fermion, color=Blue] (n4) , 
        (c1) -- [photon, thick, bend right, color=Green] (c2) ,
    };
    
\semiloop[]{c3}{c2}{-90}[][right];
\semiloop[fermion]{c2}{c3}{90};

\node[dot] at (c1);
\node[dot] at (c2);
\node[dot, white] at ($(n1) - (0, {1.30*\QCDvertA})$);

\draw [color=gray!60, fill=gray!25, ] (n11) ellipse ({0.40*\QCDvertA} and {0.70*\QCDvertA});
\draw [color=gray!60, fill=gray!25, ] (n21) ellipse ({0.40*\QCDvertA} and {0.70*\QCDvertA});
\node at (n11) {$\Sigma$};
\node at (n21) {$\Lambda$};
\node[squaredot] at (m1);
\node[squaredot] at (m2);
\node[color=Blue] at ($(n4) + (0.25, 0)$) {\(e\)};
\node[color=Blue] at ($(n3) + (0.25, 0)$) {\(\bar{\nu}\)};

\end{feynman}
\end{tikzpicture}}

\newcommand{\Exlep}{
\begin{tikzpicture}[
dot/.style={draw,circle,minimum size=2.5pt,inner sep=0pt,outer sep=0pt,fill=Green, color=Green}
]

\begin{feynman}
\vertex (n1) ;
\vertex[below=0.32*\QCDvertA of n1] (n11)    ;
\vertex[below=0.64*\QCDvertA of n1] (n12)    ;
\vertex[right=\QCDlen of n1] (n2)   ;
\vertex[right=\QCDlen of n11] (n21) ;
\vertex[right=\QCDlen of n12] (n22) ;

\vertex at ($(n1)!0.95!(n2) + (0, \QCDvertB)$) (n3) ;
\vertex at ($(n1)!0.95!(n2) + (0, \QCDvertC)$) (n4) ;

\vertex at ($(n1)!0.50!(n2)$) (m1);
\vertex[above=1.2*\QCDvertA of m1] (m2);

\vertex[dot] at ($(m1)!0.25!(n2)$) (c1);
\vertex[dot] at ($(m2)!0.80!(n4)$) (c2);

    \diagram*{
        (n1)  -- [fermion, color=Red   ] (m1)  ,
        (n11) -- [fermion, color=Red   ] (n21)  , 
        (m1)  -- [fermion, color=Apricot] (n2)  ,
        (n12) -- [fermion, color=black  ] (n22)  , 
        (n3) -- [fermion, color=Blue] (m2) , 
        (m2) -- [fermion, color=Blue] (n4) , 
        (c1) -- [photon, thick, bend left, color=Green] (c2) ,
    };

\node[dot] at (c1);
\node[dot] at (c2);
\node[dot, white] at ($(n1) - (0, {1.30*\QCDvertA})$);

\draw [color=gray!60, fill=gray!25, ] (n11) ellipse ({0.40*\QCDvertA} and {{0.70*\QCDvertA}});;
\draw [color=gray!60, fill=gray!25, ] (n21) ellipse ({0.40*\QCDvertA} and {0.70*\QCDvertA});
\node at (n11) {$\Sigma$};
\node at (n21) {$\Lambda$};
\node[squaredot] at (m1);
\node[squaredot] at (m2);
\node[color=Blue] at ($(n4) + (0.25, 0)$) {\(e\)};
\node[color=Blue] at ($(n3) + (0.25, 0)$) {\(\bar{\nu}\)};

\end{feynman}
\end{tikzpicture}
}

\title{Looking at radiative corrections to the $\Sigma^\pm{\to}\Lambda e^\pm\bar\nu$ semileptonic decay-rate ratio}

\author[a,b]{Constantia Alexandrou}
\author[a]{Simone Bacchio}
\author*[b]{Antonio Evangelista}
\author[a,b]{~~~~~~~~Andreas Konstantinou}
\author[c]{Nazario Tantalo}

\affiliation[a]{Computation-based Science and Technology Research Center, The Cyprus Institute, 20 Kavafi Str., Nicosia 2121, Cyprus}
\affiliation[b]{Department of Physics, University of Cyprus, P.O. Box 20537, 1678 Nicosia, Cyprus}
\affiliation[c]{Department of Physics, University of Tor Vergata, Rome, Italy}

\emailAdd{evangelista.antonio@ucy.ac.cy}

\abstract{We investigate the impact of radiative corrections on the ratio of the semileptonic decay rates $\Sigma^\pm{\to}\Lambda e^\pm\bar\nu$.  Phenomenological analyses point  to sizable radiative effects on  the experimentally measured ratio. From a theoretical perspective, this observable is particularly appealing because several hadronic and electroweak contributions partially cancel in the ratio, providing a cleaner probe of electromagnetic effects.  We identify the dominant contributions to the radiative corrections and develop the framework for a dedicated calculation. This represents a first step toward a precise theoretical description of baryon semileptonic decays  helping to assess their potential as precision tests of the Standard Model.}

\FullConference{
The 43rd International Symposium on Lattice Field Theory (Lattice 2026)
}

\begin{document}

\maketitle

\section{Introduction}

As experimental precision continues to improve, the reliable control of radiative and strong isospin-breaking corrections is becoming increasingly important in precision hadron physics. These corrections are commonly separated into universal long-distance effects associated with external charged particles and structure-dependent contributions that probe the internal dynamics of hadrons. The latter can modify effective couplings, decay rates, and kinematic distributions in ways that cannot be captured by point-like descriptions alone. Lattice QCD is uniquely positioned to compute these structure-dependent contributions from first principles.
Some of these effects have already been addressed on the lattice: notable examples include the $\gamma W$-box contribution to neutron $\beta$ decay~\cite{Ma:2021azh,Yoo:2023gln,Ma:2023kfr} and the structure-dependent electromagnetic corrections to leptonic decays~\cite{Carrasco:2015xwa,Giusti:2017dwk,DiCarlo:2019thl,Desiderio:2020oej,Boyle:2022lsi,Boyle:2025uuh}. These calculations demonstrate the potential of lattice methods to provide systematically improvable, nonperturbative input for precision tests of the Standard Model.

In this work, we introduce an exploratory study on the radiative corrections to the ratio of the semileptonic $\Sigma^\pm{\to}\Lambda e^\pm\bar{\nu}$ decay rates. This ratio is particularly interesting from two complementary perspectives. Phenomenologically, it provides a clean probe of isospin-breaking and electromagnetic effects in hyperon semileptonic decays. Theoretically, the partially charge-conjugate structure of the two channels leads to particularly simple infrared properties. In the isospin-symmetric limit, the corresponding QCD matrix elements are related by isospin symmetry. The Ward--Takahashi identities constrain the soft-photon contribution leading to a  factorization into the non-radiative matrix element and a universal external-state factor. As a consequence, the infrared-singular contributions to the two channels are identical and cancel in their ratio. This feature is particularly advantageous for lattice calculations since it provides a clean framework in which the remaining finite and structure-dependent contributions can be isolated and computed nonperturbatively.

\section{Phenomenological analysis}\label{sec:pheno}

The $\Sigma^+$ and $\Sigma^-$ hyperons, with quark content $uus$ and $dds$, respectively, decay semileptonically into the $\Lambda$ hyperon, with quark content $uds$, through an $u{\to}d$ or $d{\to}u$ transitions. These decays are particularly interesting because they provide the only other experimentally accessible $u{\leftrightarrow}d$ transition in the baryon sector beyond the neutron semileptonic decay. Other transitions between isospin multiplets are strongly suppressed by  small phase space, whereas the $\Sigma{\to}\Lambda$ transition is kinematically favored by the approximately $75\,\mathrm{MeV}$ mass splitting between the $\Sigma$ and $\Lambda$, which arises from their different isospin.
Nevertheless, the available phase space remains relatively small, allowing the decay rate to be organized as a perturbative expansion in the parameter $\delta_\pm$, namely  
\begin{equation}\label{eq:cabibbo}
\Gamma_{\Sigma^\pm{\to}\Lambda}^{\,\QCD}= \frac{G_\smaller{\rm F}^2\, |U_{ud}|^2  (M_{\Sigma^{\pm}}^{\EXP}\delta_{\pm})^5}{60 \pi^3}\bigg(1-\frac{3}{2}\delta_{\pm}\bigg)\bigg(f_{1}^2+3g_{1}^2\bigg)\bigg(1+\mathcal{O}(\delta_{\pm}^2)\bigg)\qq{with}\delta_\pm=\frac{M_{\Sigma^\pm}^{\EXP}-M_\Lambda^{\EXP}}{M_{\Sigma^\pm}^{\EXP}}
\end{equation}
where $f_{1}$ and $g_{1}$ denote the vector and axial-vector form factors at zero momentum transfer, respectively. In QCD, these form factors are identical for the two transitions up to strong isospin-breaking effects in the up- and down-quark masses. It is more appropriate to treat the latter together with radiative corrections when considering the full process in QCD+QED, and thus we define
\begin{equation}
    \Delta^{\!\QCD}_{\IB,\pm}\equiv\frac{\Gamma_{\Sigma^\pm{\to}\Lambda e^\pm\bar{\nu}}^{\,\QCED}}{\Gamma_{\Sigma^\pm{\to}\Lambda}^{\,\QCD}}-1\qquad\Longrightarrow\qquad\Gamma_{\Sigma^\pm{\to}\Lambda e^\pm\bar{\nu}}^{\,\QCED}= \Gamma_{\Sigma^\pm{\to}\Lambda}^{\,\QCD}\left(1+\Delta^{\!\QCD}_{\IB,\pm}\right)
\end{equation}
The content of the corrections included in $\Delta^{\!\QCD}_{\IB,\pm}$ is the focus of the following sections. Here, instead, we consider the ratio of the two decay rates, for which we obtain
\begin{equation}\label{eq:ratio_NLO}
\mathcal{R} \equiv \frac{\Gamma_{\Sigma^-{\to}\Lambda e^-\bar{\nu}}^{\,\QCED}}{\Gamma_{\Sigma^+{\to}\Lambda e^+\nu}^{\,\QCED}}
= \bigg(\frac{M_{\Sigma^-}^{\EXP}-M_\Lambda^{\EXP}}{M_{\Sigma^+}^{\EXP}-M_\Lambda^{\EXP}}\bigg)^{\!5}\!\bigg(\frac{2-3\delta_{-}}{2-3\delta_{+}}\bigg)\bigg(1+\Delta^{\!\QCD}_{\IB,-}-\Delta^{\!\QCD}_{\IB,+}+\mathcal{O}(\delta_{-}^2\!{-}\delta_+^2,\, \alpha_\EM^2,\,\alpha^\ph_\EM\delta_\pm^2)\bigg)\,.
\end{equation}
Taking the ratio of the two decay rates thus leads to substantial cancellations. In particular, the Fermi constant, CKM matrix element, and hadronic form factors cancel. The remaining corrections beyond those displayed in Eq.~\eqref{eq:ratio_NLO} are also subject to significant cancellations. In particular, the pure kinematic corrections of $\mathcal{O}(\delta_\pm^2)$ depend, in the QCD limit, on the same form factors for the two partially charge-conjugate channels. Their contribution to the ratio therefore enters only through the difference $\mathcal{O}(\delta_-^2{-}\delta_+^2)$, which is approximately $0.08\%$, thus negligible at this level of precision.

Therefore, Eq~\eqref{eq:ratio_NLO} can be separated into two conceptually distinct contributions. The first is independent of the hadronic form factors and arises entirely from the difference in phase space between the two processes, induced by the $\Sigma^\pm$ mass splitting. Despite being purely kinematic, this contribution is far from negligible, since the available energy enters the decay rate to the fifth power. Numerically, it gives a factor of approximately $1.67$, corresponding to a $67\%$ deviation from unity. The second contribution contains the isospin-breaking corrections. Comparing now this phenomenological prediction, $\mathcal{R}_{\rm phen}$, with the experimentally measured ratio, $\mathcal{R}_{\rm PDG}$~\cite{ParticleDataGroup:2026mpi}, we obtain
\begin{equation}\label{eq:Rsigma}
    \mathcal{R}_\smaller{\rm PDG}=1.35(27) ~~~\text{and}~~~ \mathcal{R}_{\rm phen}=1.665(8)(4)\cdot\left(1-\Delta^{\!\mathcal{R}}_{\IB}\right)~~~\Longrightarrow~~~ \Delta^{\!\mathcal{R}}_{\IB}\equiv\Delta^{\!\QCD}_{\IB,+}{-}\Delta^{\!\QCD}_{\IB,-}\approx 19(17)\%,
\end{equation}
where the two uncertainties in $\mathcal{R}_{\rm phen}$ originate, respectively, from the uncertainties in the baryon masses and from the estimated size of higher-order corrections. Thus, comparing the phenomenological prediction with the experimental result indicates that isospin-breaking corrections could be of order $20\%$, albeit with a large uncertainty. This motivates their first-principles study.

\section{The point-like description}\label{sec:point-like}

\begin{figure}[ht]
\centering
\vspace{-1em}
\subfloat[\label{fig:SgSe}]{\includegraphics[width=0.18\linewidth]{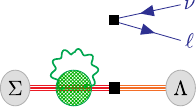}} \hfill
\subfloat[\label{fig:LmSe}]{\includegraphics[width=0.18\linewidth]{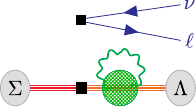}} \hfill
\subfloat[\label{fig:SgLp}]{\includegraphics[width=0.18\linewidth]{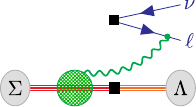}} \hfill
\subfloat[\label{fig:SgLm}]{\includegraphics[width=0.18\linewidth]{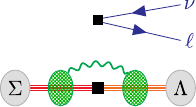}} \hfill
\subfloat[\label{fig:LmLp}]{\includegraphics[width=0.18\linewidth]{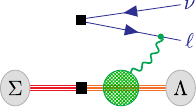}}
\\
\subfloat[\label{fig:SgQm}]{\includegraphics[width=0.18\linewidth]{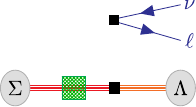}} \hfill
\subfloat[\label{fig:LmQm}]{\includegraphics[width=0.18\linewidth]{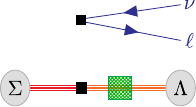}} \hfill
\subfloat[\label{fig:SgRe}]{\includegraphics[width=0.18\linewidth]{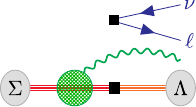}} \hfill
\subfloat[\label{fig:LmRe}]{\includegraphics[width=0.18\linewidth]{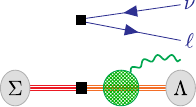}} \hfill
\subfloat[\label{fig:LpRe}]{\includegraphics[width=0.18\linewidth]{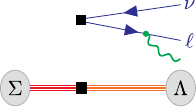}}
\caption{The classes of virtual radiative (a)--(e), strong isospin-breaking (f)--(g) and real photon emission (h)--(j) correction diagrams contributing to the decay rate of Eq.~\eqref{eq:cabibbo}.}
\label{fig:point-like}
\vspace{-0.2cm}
\end{figure}

In this section, we introduce the point-like approximation with the main goal of discussing the cancellation of infrared (IR) divergences. In the soft-photon limit, hadrons can be treated as point-like particles and IR divergences naturally arise from any radiative contribution. A general feature of radiative corrections is that such divergences cancel order-by-order when virtual photons, diagrams (a)--(e) in Fig.~\ref{fig:point-like}, and real photon emissions, diagrams (h)--(j), are combined together, as they appear equal but with opposite sign in the two contributions. This is known as the Bloch--Nordsieck mechanism~\cite{Bloch:1937pw,Yennie:1961ad}.
A convenient feature of the ratio in Eq.~\eqref{eq:ratio_NLO} is remarkably that IR-divergent terms cancel also diagram-by-diagram, thanks to the partially charge-conjugate structure of the $\Sigma^\pm {\to} \Lambda e^\pm \nu$ transitions. This allows the finite structure-dependent contribution to be isolated without introducing real-photon emission, in close analogy with lattice calculations of electromagnetic hadron mass splittings~\cite{Giusti:2017dmp}. Other diagrams depicted in Fig.~\ref{fig:point-like} are the strong isosospinbreaking corrections due to the up--down mass splitting and depicted in digrams (f)--(g). Also these are important to be threated together with the radiative correction because they cancel UV divergences of the self-energy diagrams (a)--(b).

\subsection{Cancellation of IR divergences through Ward--Takahashi identities}
\label{sec:ward}

A particularly transparent way to see this cancellation is provided by the Ward--Takahashi identity. We consider the non-radiative matrix element entering the semileptonic decay
\begin{equation}
\mathcal{M}_\W^\nu =
\langle \Lambda(p')|J_\W^\nu(0)|\Sigma(p)\rangle,
\end{equation}
and the corresponding time-ordered product with one electromagnetic current,
\begin{equation}
T_{\gamma W}^{\mu\nu}(k) =
\int d^4x\,e^{ik\cdot x}
\langle \Lambda(p')|
T\!\left\{J_{\EM}^\mu(x)J_\W^\nu(0)\right\}
|\Sigma(p)\rangle.
\end{equation}
The Ward--Takahashi identity relates the contraction of the electromagnetic current with the time-ordered product of the commutator of the electromagnetic charge with the weak current,
\begin{equation}
k_\mu T_{\gamma W}^{\mu\nu}(k)
=
i\langle \Lambda(p')|
[Q^\ph_{\EM},J_\W^\nu(0)]
|\Sigma(p)\rangle
=
i\left(Q^\ph_\Lambda-Q^\ph_\Sigma\right)\mathcal{M}_\W^\nu
=
-iQ^\ph_\Sigma\mathcal{M}_\W^\nu,
\end{equation}
where $Q$ is the electric charge of the states and we have used $Q^\ph_\Lambda{=}0$ since $\Lambda$ is neutrally charged.
To make the infrared structure explicit, we consider the soft-photon limit, $k{\to}0$. In this limit, the singular part of the time-ordered product is determined by the coupling of the photon to the external charged states. The leading behavior is therefore
\begin{equation}
T_{\gamma W}^{\mu\nu}(k)
=
-iQ^\ph_\Sigma\frac{p^\mu}{p\cdot k}\mathcal{M}_\W^\nu
+
T_{\gamma W,\mathrm{fin}}^{\mu\nu}(k)
\quad\Longrightarrow\quad
k_\mu T_{\gamma W, \mathrm{fin}}^{\mu\nu}(k)
=
0\,,
\end{equation}
where $T_{\gamma W,\mathrm{fin}}^{\mu\nu}(k)$ is finite in the soft-photon limit and is completely perpendicular to the photon momenta. 
Since we compute radiative corrections at leading order, the matrix element $\mathcal{M}_\W^\nu$ is taken in the isoQCD limit and is equal for both channels. Thus, the IR term in $T_{\gamma W}^{\mu\nu}$ is also identical, apart from the sign associated with the charge $Q^\ph_{\Sigma}$ of the $\Sigma^\pm$ hyperons.

This sign is then compensated by the charge of the other particle to which the electromagnetic current couples. As illustrated in Fig.~\ref{fig:point-like}, the second electromagnetic vertex can involve the $\Sigma$ itself [diagram (a)], the charged lepton [diagram (c)], or the $\Lambda$ hyperon [diagram (d)]. The latter, as well as any other contributions involving the $\Lambda$ [diagrams (d)--(e)], vanish because $Q^\ph_\Lambda{=}0$.
For the remaining contributions, the IR part of the virtual-photon correction, $\mathcal{M}_{\rm virt}^{\IR}$, has, in the soft-photon limit, the schematic form
\begin{equation}
\mathcal{M}_{\rm virt}^{\IR}
=
e^2\mathcal{M}^\ph_\W
\left[
Q_\Sigma^2\,I^\ph_{\Sigma\Sigma}
+
Q_\ell^2\,I^\ph_{\ell\ell}
+
Q^\ph_\Sigma Q^\ph_\ell\,I^\ph_{\Sigma\ell}
\right]
\,,
\end{equation}
where the functions $I^\ph_{\Sigma\Sigma}$, $I^\ph_{\ell\ell}$, and $I^\ph_{\Sigma\ell}$ contain the infrared-divergent loop integrals and depend only on the momenta and masses of the external particles. Since
\begin{equation}
Q_{\Sigma^+}^2=Q_{\Sigma^-}^2\,,\quad\;
Q_{e^+}^2=Q_{e^-}^2\,,\quad\;
Q^\ph_{\Sigma^+}Q^\ph_{e^+}=Q^\ph_{\Sigma^-}Q^\ph_{e^-}
\quad\Longrightarrow\quad \left.\mathcal{M}_{\rm virt}^{\IR}\right|_{\Sigma^+}
=
\left.\mathcal{M}_{\rm virt}^{\IR}\right|_{\Sigma^-}\,,
\end{equation}
and the corresponding non-radiative matrix elements are equal in the isoQCD limit, we obtain that the IR divergences are equal.
Thus, although the soft-photon pole in the $\gamma W$ amplitude changes sign between the two partially charge-conjugate channels, the virtual correction contains two electromagnetic couplings, such that the complete infrared-divergent contribution is even between the two channels. The IR divergences therefore cancel term-by-term in the ratio of the two decay rates, both within virtual as well as real radiative corrections. We can consequently write
\begin{equation}\label{eq:RC}
    \frac{\Gamma_{\Sigma^-{\to}\Lambda e^-\bar{\nu}}^{\,\QCED}}{\Gamma_{\Sigma^+{\to}\Lambda e^+\nu}^{\,\QCED}}
    \approx
    \frac{\Gamma_{\Sigma^-{\to}\Lambda }^{\,\QCD}}{\Gamma_{\Sigma^+{\to}\Lambda}^{\,\QCD}}
    \bigg[
    1+\underbrace{\alpha_{\EM}
    \left(
    C^{\,\smaller{\rm RC\,finite}}_{\Sigma^-{\to}\Lambda e^-\bar{\nu}}
    -
    C^{\,\smaller{\rm RC\,finite}}_{\Sigma^+{\to}\Lambda e^+\nu}
    \right)+\Delta m_{ud} \left(
    C^{\,\smaller{\rm SIB}}_{\Sigma^-{\to}\Lambda}
    -
    C^{\,\smaller{\rm SIB}}_{\Sigma^+{\to}\Lambda}
    \right)}_{-\Delta^{\!\mathcal{R}}_{\IB}}
    \bigg],\vspace{-0.3cm}
\end{equation}
up to higher orders in $\alpha_\EM$ and $\Delta m_{ud} {=} m_u{-}m_d$. The quantities $C^{\,\smaller{\rm RC\,finite}}$ denote the finite parts of the leading-order radiative corrections after the infrared-divergent contributions are removed and $C^{\,\smaller{\rm SIB}}$ are strong isospin-breaking corrections to the form factors.

\subsection{Cancellation of radiative corrections in the point-like limit}
\label{sec:point-like}

The expression in Eq.~\eqref{eq:RC} leads us to another important observation: any radiative correction that is even between the two channels cancels at leading order.
From the above discussion, all virtual radiative corrections of Fig.~\ref{fig:point-like} cancel in the point-like limit, since they are either even under partial charge conjugation or vanish because the $\Lambda$ is electrically neutral.
Concerning the real-photon contributions, we need to consider the squared amplitude, in which diagrams (h)--(j) combine with each other to form the corresponding electromagnetic correction. Here too, the resulting charge dependence appears through either charge-even or (partially-)neutral-charge combinations. The real-photon contributions are thus also subject to substantial cancellation in the ratio. An important caveat concerns possible differences in the experimental treatment of the real photon emission, such as the sensitivity to, and threshold between, hard- and soft-photon emissions. Such differences can thus survive but, however, be computed analytically within the point-like approximation.

The remaining corrections to consider are those due to strong isospin-breaking effects induced by the up--down quark mass difference $\Delta m_{ud}$, corresponding to diagrams (f)--(g) of Fig.~\ref{fig:point-like}.
In this case, the strong isospin-breaking corrections associated with the $\Lambda$ hyperon [diagram (g)] are identical for the two channels and therefore cancel in the ratio. On the other hand, when these corrections affect the initial $\Sigma$ state [diagram (f)], they differ between the two processes and thus survive in the ratio. Their leading effect is a modification of the $\Sigma^\pm$ masses, and hence of the available phase space, as discussed in the phenomenological analysis of Sec.~\ref{sec:pheno}. However, they also affect the corresponding hadronic form factors, since, in the presence of SU(2) isospin breaking, $\Sigma^+$ and $\Sigma^-$ are distinct states and thus produce characteristic matrix elements.
Let us consider these effects explicitly. The vector form factor $f_1$ is constrained by conserved-vector-current (CVC) relations and by the Ademollo--Gatto theorem~\cite{Ademollo:1964sr}. In particular, since the SU(3)-symmetric value for the $\Sigma^\pm{\to}\Lambda$ transition is $f_1^{\rm SU(3)}=0$, the leading symmetry-breaking contributions are of $\order{\delta_\pm^2}$. Thus, $f_{1_\pm}=c_{\pm}\delta_{\pm}^2$, and since $f_1$ enters the decay rate quadratically, its contribution is of $\order{\delta_\pm^4}$ and can therefore be neglected at the order considered here. We are thus left with strong isospin-breaking effects in the axial-vector form factor $g_1$, which can be parametrised as
\begin{equation}
g_{1, \pm} = g_1\left(1 + \frac{\Delta g_{\pm}}{g_1}\right)
\quad \Longrightarrow \quad
\frac{3g_{1, -}^2}{3g_{1, +}^2} \simeq
1 - 2\,\dfrac{\Delta g_{+} - \Delta g_{-}}{g_1}
\quad \Longrightarrow \quad
\frac{\Delta^\mathcal{R}_{\IB}}{2} = \dfrac{\Delta g_{+} - \Delta g_{-}}{g_1}
\,.
\end{equation}
In conclusion, the point-like approximation, combined with the comparison between the phenomenological prediction and experiment in Eq.~\eqref{eq:Rsigma}, suggests that strong isospin-breaking effects in the axial charge are of order 10(8)\%, unexpectedly large.

\section{Radiative corrections in QCD+QED}

\begin{figure}[t]
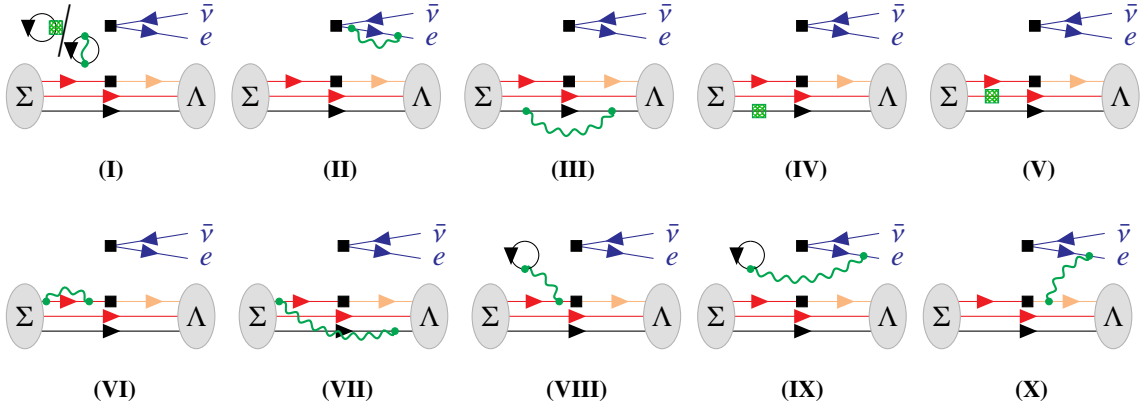

\centering
\renewcommand{\thesubfigure}{\Roman{subfigure}}
\setlength{\QCDlen}{0.15\textwidth}
\pgfmathsetlength{\QCDvertA}{0.27*\QCDlen}
\pgfmathsetlength{\QCDvertB}{1.47*\QCDvertA}
\pgfmathsetlength{\QCDvertC}{0.93*\QCDvertA}
\vspace{-1em}
\subfloat[\label{fig:diag1}]{\Sea}
\subfloat[\label{fig:diag2}]{\Selep}
\subfloat[\label{fig:diag3}]{\Sestr}
\subfloat[\label{fig:diag4}]{\Qmstr}
\subfloat[\label{fig:diag7}]{\Qmup}
\\
\subfloat[\label{fig:diag6}]{\Seup}
\subfloat[\label{fig:diag5}]{\Exup}
\subfloat[\label{fig:diag8}]{\Seaval}
\subfloat[\label{fig:diag9}]{\Sealep}
\subfloat[\label{fig:diag10}]{\Exlep}
\caption{For each class of virtual-photon or strong-isospin-breaking correction diagrams, one representative diagram is shown. The bottom black quark line represents the strange quark, while the other red or orange quark lines represent either up or down quarks, depending on the transition.}
\label{fig:structure}
\end{figure}

The situation changes substantially once the structure dependence of the interaction is taken into account, as depicted in Fig.~\ref{fig:structure} with representative diagrams. In this case, the sea-quark contributions [diagram (I)], the lepton self-energy corrections [diagram (II)], the strange-quark self-energy corrections [diagram (III)], and the strange-quark mass shift [diagram (IV)] associated with the change from the QCD to the QCD+QED scale setting are all even and therefore cancel in the ratio at leading order. The remaining contributions are non-vanishing and can be further classified according to whether they preserve the factorisation of the hadronic and leptonic amplitudes in the decay rate:
\begin{itemize}
\item In the first case, we have up--down strong isospin-breaking effects [diagram (V)], valence up- and down-quark self-energy contributions [diagram (VI)], photon exchange between valence quarks [diagram (VII)], and photon exchange between a valence quark and a sea-quark loop [diagram (VIII)]. These contributions preserve factorisation, modifying only the hadronic matrix elements, and can therefore be absorbed into a redefinition of the form factors.
\item In the second case, we have the electromagnetic interaction of the lepton with either a sea quark [diagram (IX)] or a valence quark [diagram (X)]. These contributions give rise to the $\gamma W$-box correction, which is genuinely non-factorisable and structure-dependent, and is notoriously challenging to compute. In particular, intermediate states can generate poles that obstruct a straightforward mapping between Minkowski and Euclidean space. Current lattice approaches to neutron beta decay address this issue through a combination of lattice and dispersive methods~\cite{Ma:2021azh,Yoo:2023gln,Ma:2023kfr}. In the present case, however, the situation may be less severe, since the problematic final-state interaction, diagram (e) in Fig.~\ref{fig:point-like}, is suppressed in the point-like description by the neutral charge of the $\Lambda$ hyperon.  Nevertheless, recent developments in spectral density approaches can be used to address this problem in full generality~\cite{Patella:2024cto,DiCarlo:2026kpv}.
\end{itemize}

These contributions can be summarised by the following expression:
\begin{align}
\Delta_{\IB}^{\mathcal{R}} & 
= \Delta m_{ud}\Big(M_u^{+} - M_d^{+}\Big) + \Big(e_u^2 - e_d^2\Big)\Big(V_{uu}^{+} - V_{dd}^{+}\Big) + e_s\Big(e_u - e_d\Big)\Big(V_{us}^{+} - V_{ds}^{+}\Big)  
\label{eq:structure}\\
& 
+ \Big(e_u^2 - e_d^2\Big)\Big(U_{u\ell}^{+} - U_{d\ell}^{+}\Big) + e_s \Big(e_u - e_d\Big)\Big(U_{us}^{+} - U_{ds}^{+}\Big) + e_c\Big(e_u - e_d\Big)\Big(U_{uc}^{+} - U_{dc}^{+}\Big) 
\nonumber \\
& 
+ e \Big(e_u + e_d\Big)\Big(L_{ue}^{+} + L_{de}^{+}\Big) + 2 e e_s L_{se}^{+} + 2ee_c L_{ce}^{+}\,.
\nonumber
\end{align}
Here, we have exploited the exchange symmetry $u {\leftrightarrow} d$ and $\Sigma^{+} {\leftrightarrow} \Sigma^{-}$ of matrix elements evaluated in the SU(2) limit, and expressed all corrections in terms of the $\Sigma^{+}$ decay, corresponding to an $uus{\to}uds$ transition. The purpose of this equation is to highlight the charge factors and the nature of the contributions entering the isospin-breaking corrections. These can originate either from the $u$- or $d$-quark mass dependence, $M_{u\!/\!d}$; from electromagnetic interactions between valence quarks of flavours $q$ and $q'$, $V_{qq'}$; from electro-unquenching contributions, in which a valence quark of flavour $q$ interacts electromagnetically with a sea-quark loop of flavour $q'$, $U_{qq'}$, with $\ell$ denoting a $u/d$-quark loop; or from quark-lepton interactions, $L_{qe}$, which break factorization. All these contributions should, of course, be understood as corrections to the decay rate through the corresponding matrix elements. Consequently, each term contains an integral over phase space and a kinematic kernel, both computed using the experimental masses of the baryons, as in Eq.~\eqref{eq:ratio_NLO}, and momentum-dependent squared amplitudes obtained by properly contracting the leptonic and hadronic tensors. Thus, Eq.~\eqref{eq:structure} provides the structure of the isospin-breaking corrections, up to an explicit analytic definition of each contribution, which will be provided in future work.

Two additional remarks are in order. First, the electro-unquenching contributions [diagrams (VIII) and (IX)], mainly represented by $U$ but also appearing in $L$, are generally expected to be suppressed because they involve sea-quark loops, as observed for quantities in which these contributions have been computed~\cite{Borsanyi:2020mff,RC:2025zoa}. Second, real-photon emissions have deliberately not been considered here, since the available phase space is at most $75$\,MeV and thus the soft-photon limit and point-like description discussed in Sec.~\ref{sec:point-like} are sufficient, leading to the same above mentioned cancellations.

\section{Conclusions and outlook}

The key theoretical features of isospin-breaking and radiative corrections to the ratio of the semileptonic decay rates of the charged $\Sigma$ hyperons into $\Lambda$ are presented. In particular, we show that the partially charge-conjugate structure of the two decay channels leads to a substantial cancellation of the infrared-sensitive and point-like contributions,  making the ratio of their decays a particularly clean framework for isolating the remaining structure-dependent effects. We identify the different classes of contributions that survive in the ratio and discuss their relative theoretical and computational challenges.
Our immediate goal is to compute the leading contributions associated with the valence quarks using the RM123 method and ETMC ensembles. This calculation will follow the methodology that we have recently developed and validated for the computation of the scale-setting~\cite{Evangelista:2025dzi} and isospin-breaking effects in baryon mass splittings~\cite{Alexandrou:2026eko}. Beyond providing the first quantitative determination of the leading structure-dependent corrections to the $\Sigma^\pm{\to}\Lambda e^\pm\bar{\nu}$ decay ratio from lattice QCD+QED, this calculation represents an important first step towards a fully nonperturbative treatment of radiative corrections to semileptonic baryon decays.

The $\Sigma{\to}\Lambda$ transition is particularly interesting in this context because it is an isovector $u{\leftrightarrow}d$ transition closely connected to the nucleon case. In particular, previous phenomenological studies of the nucleon axial coupling $g_A$ found that radiative corrections associated with pion-cloud effects can be sizeable~\cite{Cirigliano:2022hob}. This observation provides additional motivation for investigating the size of the corresponding structure-dependent corrections in the $\Sigma{\to}\Lambda$ transition. While the phenomenological analisys does not by itself determine the origin of the correction in the hyperon decay, it highlights the possibility that the axial form factor may receive non-negligible radiative and isospin-breaking contributions. Moreover, the $\Sigma{\to}\Lambda$ transitions are expected to attract renewed experimental interest in light of the upcoming experimental programme at FAIR~\cite{Senger:2018zax}.

\setlength{\parskip}{0pt}
\section*{Acknowledgements}
We thank Matteo Di Carlo, Andrzej Kupsc and Piotr Salabura for useful discussions. C.A., S.B. A.E. and A.K. acknowledge support from the projects EXCELLENCE/0524/0459 (IMAGE-N), EXCELLENCE/0524/0017 (MuonHVP), POST-DOC/0925-ROandOther/0148  (HiQO), and POST-DOC/0524/0001 (Baryon8) implemented under the programme of social cohesion “THALIA 2021-2027” co-funded by the European Union and the Republic of Cyprus, through the Research and Innovation Foundation. S.B. also acknowledges support by the project OPPTY-ERC/0524/0006 (QWERTY), funded by the Republic of Cyprus, through the Research and Innovation Foundation. 

\bibliography{refs}
\bibliographystyle{JHEP}

\end{document}